\documentstyle[12pt]{article}
\newcommand{\tr}[1]{\,{\rm tr}\,#1\,}

\begin{document}
\def\newmathop#1{\mathop{#1}\limits}
\def\osim#1{\displaystyle\newmathop{#1}_{\lambda\to 0}}
% $A\osim{\sim}B$
\def\linfty#1{\displaystyle\newmathop{#1}_{s\to \infty}}
% $A\linfty{\sim}B$
\def\rr#1{\displaystyle\newmathop{#1}_{~~s \to \infty ,~ t~fixed~~}}
% $A\rr{\sim}B$
\title{ \begin{flushright}
{\small SMI-15-94 }
\end{flushright}
\vspace{2cm}
Anisotropic Asymptotics\\ and\\
High Energy Scattering  }
\author{I.Ya.Aref'eva \\
{\it Steklov Mathematical
Institute,}\\ {\it Vavilov st.42, GSP-1,117966, Moscow , Russia }
\\ and \\ I.V.Volovich \\
{\it Steklov Mathematical
Institute,}\\ {\it Vavilov st.42, GSP-1,117966, Moscow , Russia }
}
\date {$~$}
\maketitle
\begin {abstract}
Recently E.Verlinde and H.Verlinde have suggested an effective
two-dimensio\-nal theory describing the high-energy scattering
in QCD. In this report we attempt to clarify some issues
of this suggestion. We consider {\it anisotropic asymptotics} of correlation
functions for scalar and gauge theories in four dimensions.
Anisotropic asymptotics
describe behaviour of correlation functions when some components of
coordinates are large as compare with others components. It is
occurred that (2+2) anisotropic asymptotics for 4-points functions
are related with the well known Regge regime of scattering amplitudes.

We study an expansion of correlation functions
with respect to the rescaling parameter $\lambda$ over a part of variables
(anisotropic $\lambda$-expansion). An effective theory describing
the anisotropic limit of free scalar field contains two 2 dim conformal
theories. One of them is a conformal theory in configuration space
and another one is a conformal theory in momentum space. In some special
cases ,in particular for the Wilson line correlators in gauge theories,
 the leading term of the anisotropic expansion involves only one of the
conformal theories and it can be described by  an effective theory
 with an action being
a dimensional reduction of the original action.

\end {abstract}

\section{Introduction}
\setcounter{equation}{0}

Recently E.Verlinde and H.Verlinde  \cite {VV} have suggested
a new approach to high energy scattering in QCD.
They performed a rescaling of the
longitudinal coordinates inside the Yang-Mills action
and reduced the full theory to a two-dimensional
sigma-model in the transversal subspace. Moreover they have assumed a
special ansatz for the truncated action which was engaged to reproduce the
$\log s$ dependence of the high energy of amplitudes in QCD.

Several papers there appeared in which the approach
of  \cite {VV} have been discussed
 \cite {AI}-\cite {Rey}.
Quantum corrections to the
longitudinal dynamics in the Verlindes'
approach have been considered in  \cite {AI}
A nonperturbative approach to the Regge regime in QCD
based on an anisotropic lattice gauge theory has
been suggested in  \cite {AII}.

The necessity of non-perturbative study of
QCD in the Regge regime of large energies
$\sqrt{s}\to \infty$
and fixed momentum transfers $q$ $(q^{2}=-t)$, $|q|\sim $ $1Gev$ has
been  emphasized by  Nachtmann \cite {Nach}.
Let us recall that problems of nonperturbative investigations
of QCD have been discussed for many years.
Some hopes were laied on an analogy between four-dimension Yang-Mills
theory and two-dimensional chiral field   \cite {Pol,Ar}.
There is a well known conjecture that the long-distance dynamics of
gauge theories  in the confining phase is described in terms of
two-dimensional conformal field theory  \cite {LuSy,Gl}.
In the recent years Lipatov  \cite {Lip} has made suggestions
on a relationship between QCD at high energies and
a two-dimensional field theory (see also
\cite {KL} for further developments). This approach is based on the
solution of the unitarity condition. Recently Faddeev and Korchemsky
 \cite {FK}
have found that the Lipatov two dimensional
effective theory  is completely integrable
and they have studied it by using of a generalized Bethe ansatz.

The Verlindes' approach is based on elegant rescaling arguments
but there are some questions which deserves a further
study. In particular, it is not enough clear
why one should use an ansatz  \cite {VV} which fixes $\log s$ expansion and
which makes contact with the usual perturbative answer and reggeization.
Also there are questions about ultraviolet and infrared divergences.
Some aspects of these problems have been discussed in  \cite {VV}- \cite {AII}

In this talk  we discuss  some of these
questions. We will show that the rescaling arguments can be
understand by using the notion of  {\it anisotropic
asymptotics}.  Let us explain  what we mean by
 the anisotropic
asymptotics.
If  $x^{\mu}$ are space-time coordinates, one denotes
$x^{\mu}=(y^{\alpha}, z^{i})$, where $\alpha =0,1$, $i=2,3$ for 2+2
decomposition and  $\alpha =0$,
$i=1,2,3$ for 1+3 decomposition. We are interested in
the evaluation of asymptotics of correlation
functions for a field  $\Phi _{j}(\lambda y,z)$,
\begin {equation} %---------------------------------------------------
                                                          \label {1.1}
<\Phi _{j_{1}}(\lambda y_{1},z_{1})\Phi _{j_{2}}(\lambda y_{2},z_{2})...
\Phi _{j_{n}}(\lambda y_{n},z_{n})>
\end   {equation} %---------------------------------------------------
when $\lambda\to 0$ (or $\lambda\to \infty$), i.e.
when some of variables are much larger than others.
If one rescales all the variables then one deals with
 the usual short or large
distance behaviour of the theory. These
 {\it isotropic} asymptotics are given
by the factors defined by ultraviolet or infrared dimensions of the
corresponding Green function multiplied on corresponding anomalous dimensions
defined  by the standard
renormalization group approach  \cite {BogS,Coll}.
The leading terms of the anisotropic asymptotics can be expected to be given
by anisotropic dimension analysis and be related with
 {\it anisotropic operator
product expansion }(AOPE)

\begin {equation} %---------------------------------------------------
                                                          \label {1.2}
\Phi _{i_{1}}(\lambda y_{1},z_{1})\Phi _{i_{2}}(\lambda y_{2},z_{2})
\osim{\sim}
\sum C_{i_{1}i_{2}}^{n_{1}n_{2}}(y_{1},y_{2},\lambda)
{\cal O}_{n_{1}n_{2}}(y_{1},z_{1},z_{2})
\end {equation} %---------------------------------------------------

 Matveev, Muradyan and Tavkhelidze \cite {MMT} have used a generalized
dimensional
analysis
 to   derive an automodel behaviour of
differentional cross-sections for hadrons-hadrons reactions.

 Our approach to anisotropic asymptotics was  stimulated by
 recent consideration of stochastic limit in QFT \cite {ALV} where in fact the
 $(1+3)$ anisotropic asymptotics
has been considered, i.e. an asymptotic behaviour
of quantum field $\Phi (\lambda t,x)$ after
time rescaling was evaluated.

There is an analogy between the anisotropic $(1,d-1) $ $\lambda$ rescaling
and the high temperature limit of lattice gauge theory \cite {CDP}.

 The anisotropic rescaling
gives generally a more complicated effective action
as compare with isotopic rescaling
since after making a change of variables in the functional integral
one gets an anisotropic action with the small parameter $\lambda$
in front of some terms  with derivatives.
 However some simplifications take place for the case of massless
gauge theory.
An essential difference between scalar and vectors theories
is due to that in the latter case the anisotropic rescaling acts
different on different
components and just in the zero order on rescaling parameter we get  a
nontrivial action.
In this case one deals in  some sense  with  a dimensional
reduction. However the naive rescaling arguments do not always work
because we are working with distributions.
In particular,
the naive derivation of the effective action cannot be used
even for the free scalar field.

As we will see the anisotropic asymptotics corresponding to the (2+2)
decomposition of 4-point correlation function is related with behaviour of
scattering amplitudes in the Regge regime (s large, t is fixed).
In the standard study of the Regge regime in QFT \cite{Wu,LipR}
 one first expands
the functional integral in the perturbation series with respect
to the coupling constant and then calculates the limit $s\to \infty$ for $t$
fixed  for each
diagram and finally sum up the leading terms. It is rather remarkable that
in the perturbation theory the leading terms exhibit the following structure
\begin {equation} %---------------------------------------------------
                                                          \label {1.3}
A(s,t)\rr{\sim}\sum g^{4n} (\ln s)^{n}I_{n}(t)
\end   {equation} %---------------------------------------------------
where $I_{n}(t)$ can be represented by two-dimensional Feynman diagrams in the
transversal  space ($z-$space). This representation takes place for all
theories and it is
interesting to understand an origin of such representation without
doing an examination of individual diagrams.
We would like to emphasize that  we are going
to perform an expansion with respect to
$\lambda$ in the functional integral, i.e. this approach
is a non-perturbative one. After the evaluation
the leading term with respect to  $\lambda$ one can try to
take into account next terms in the $\lambda$-expansion.
So it seems that the $\lambda$-expansion looks like
a systematic method similar to the  semiclassical
expansion or to the renormalization group approach.
Apparently it should exist a renormalization group
approach corresponding to the anisotropic rescaling and the corresponding
Callan-Symanzik equations could give,
in particular, the Regge form of the amplitude
 (see  \cite {Arb}).

\section{
Asymptotics of correlation functions for scalar theories}
\subsection{Isotropic asymptotics of correlation functions}
\setcounter{equation}{0}

It is well known that  asymptotics of perturbative connected correlations
functions in a local field theory for small  coordinates are
governed by  the ultraviolet dimension $\mbox{dim}_{n,k}$  of the
corresponding diagram
\begin {equation} %---------------------------------------------------
                                                          \label {UA}
<\phi(\lambda x_{1}) \phi(\lambda x_{2})~... ~
\phi(\lambda x_{n})>^{(k)} \osim{\sim}~
(\lambda  )^{-\mbox{dim}_{n,k}}G^{(k)}(x_{1},x_{2},...x_{n}).
\end   {equation} %---------------------------------------------------
Here $k$ means the k-order of the perturbative theory.

The asymptotics
(\ref {UA}) follows from the dimensional analysis. Indeed,
considering for definiteness the case of
selfinteracting scalar  field $\phi $  in  d-dimensional space-time
$$<\phi(\lambda x_{1}) \phi(\lambda x_{2})~... ~
\phi(\lambda x_{n})>=
$$
\begin {equation} %---------------------------------------------------
                                                          \label {sa}
\int \phi (\lambda x_{1})...\phi (\lambda x_{n})
\exp \{ i\int d^{d}x[\frac{1}{2}(\partial \phi)^{2}-\frac{m^{2}}{2}\phi ^{2}-
V(\phi ) ]\}
d\phi
\end   {equation} %---------------------------------------------------
and performing the change of variables
\begin {equation} %---------------------------------------------------
                                                          \label {scv}
\phi (\lambda x)=(\lambda  )^{\frac{2-d}{2}}
\tilde {\phi }(x)
\end   {equation} %---------------------------------------------------
in the path integral and also the change of variables $x\to \lambda x$
in the action one gets
$$<\phi (\lambda x_{1})...\phi(\lambda x_{n})>=
(\lambda  )^{n\frac{2-d}{2}}
\int \tilde {\phi }(x_{1})...\tilde {\phi }(x_{n})\cdot
$$
\begin {equation} %---------------------------------------------------
                                                          \label {sar}
\exp \{ i\int d^{d}x[\frac{1}{2}(\partial \tilde {\phi})^{2}
-\lambda ^{2}\frac{m^{2}}{2}\tilde {\phi}^{2}
-\lambda  ^{d}V(\lambda  ^{\frac{2-d}{2}}\tilde {\phi} )
] \}d\tilde {\phi}.
\end   {equation} %---------------------------------------------------

To get the ultraviolet asymptotic, i.e. $\lambda  \to 0$ one can neglect
the mass term and we see that the ultraviolet behaviour is given by the UV
index of the corresponding diagram.

In particular, for $d=4$, $V(\phi)=g\phi ^{4}$  (\ref {sar}) gives
\begin {equation} %---------------------------------------------------
                                                          \label {su}
<\phi (\lambda x_{1})...\phi(\lambda x_{n})>\osim
{\sim}
{}~(\lambda  )^{-n} <\phi (x_{1})...\phi (x_{n})>_{m=0}
\end   {equation} %---------------------------------------------------

Ultraviolet divergences do not destroy this formal estimation  at least
for renormalizable theories and (\ref {UA}) takes place after suitable
renormalizations. In this case using the renormalization
group equations one can get
the logarithmic corrections to the formula (\ref {UA}). Let us assume
that we are working in the BPHZ subtraction scheme with a
subtraction point $\mu$ then from the dimensional analysis  it follows
\begin {equation} %---------------------------------------------------
                                                          \label {UAC}
\tilde {G}_{\mbox{ren}}(\{\frac{p_{i}}{\lambda} \},g,\mu) =
(\frac{1}{\lambda} )^{dim_{n,k}}\tilde {G}_{\mbox{ren}}(\{p_{i}\},g,
\lambda\mu),
\end   {equation} %---------------------------------------------------
Here  $\tilde {G}_{\mbox{ren}}$  is the Fourier transformation of the
renormalized Green function $G _{\mbox{ren}}$ and asymptotics for small $x$
are related with asymptotics of $\tilde {G}_{\mbox{ren}}$  for large momenta,
$$<\phi(\lambda x_{1}) \phi(\lambda x_{2})~... ~
\phi(\lambda x_{n})>_{ren,\mu} =
$$
\begin {equation} %---------------------------------------------------
                                                          \label {FT}
(\lambda )^{-d(n-1)}\int e^{i\sum p_{i}x_{i}}
\delta ^{d}(\sum p_{i})\tilde{G}_{ren}(\{\frac{p_{i}}{\lambda } \},g,\mu)
\prod dp_{i}.
\end   {equation} %---------------------------------------------------
In the l.h.s. of (\ref {UAC}) one has now the renormalized Green functions
but with the subtraction point different from the initial one.
To restore the
subtraction point one can use the renormalization group invariance and
compensate
the  shift of the subtraction point by the change of the
coupling constant and the
renormalization of the  wave function
\begin {equation} %---------------------------------------------------
                                                          \label {UAR}
\tilde{G}_{\mbox{ren}} (\{p_{i}\},g, \lambda\mu) =
\xi (\frac{\mu}{\lambda} )^{-k/2}\tilde{G}_{\mbox{ren}} (\{p_{i}\},
g(\frac{\mu}{\lambda}),\mu)
\end   {equation} %---------------------------------------------------
here $\xi$ is anomalous dimension.
One finally gets the following well-known formula
for the Fourier transformation
of the correlation functions
\begin {equation} %---------------------------------------------------
                                                          \label {su1}
\tilde{G}_{\mbox{ren}}(\{\frac{ p_{i}}{\lambda}\},g, \mu) =
\xi (\frac{\mu}{\lambda})^{-k/2}(\frac{1}{\lambda} )^{\mbox{dim}_{k,n}}
\tilde{G}_{\mbox{ren}}(\{p_{i}\},
g(\frac{\mu}{\lambda}),\mu)
\end   {equation} %---------------------------------------------------

\subsection{(2+2)-anisotropic
asymptotics for scalar theories}

Let us study the behaviour of correlation
functions when only some components of coordinates in a preferable frame
are
supposed to be large or small. One of the most important examples
corresponds to the (2+2)-decomposition and it describes the case
when all longitudinal components in the central mass frame are assumed
much smaller then transersal ones.
Let $x^{\mu}$ be coordinates in the 4-dimensional Minkowski space-time
and  denote
$x^{\mu}=(y^{\alpha},z^{i})$,
$\alpha =0,1$, $i=2,3$.
 The 2+2 anisotropic asymptotics
of  Green functions of scalar selfinteracting theory
describes the behaviour of
the following correlation functions
\begin {equation} %---------------------------------------------------
                                                          \label {as1}
G_{n}(\{\lambda y_{i},z_{i}\})=<\phi (\lambda y_{1},{z}_{1})
\phi (\lambda y_{2},{z}_{2})...
\phi (\lambda y_{n},{z}_{n})>
\end   {equation} %---------------------------------------------------
for $\lambda \to 0$.

For the scalar self interacting theory
the correlation function (\ref {as1}) in the Euclidean
 regime is given by
$$<\phi (\lambda y_{1},{z}_{1})
\phi (\lambda y_{2},{z}_{2})...
\phi (\lambda y_{n},{z}_{n})>=\int \phi (\lambda y_{1},{z}_{1})
\phi (\lambda y_{2},{z}_{2})...
\phi (\lambda y_{n},{z}_{n})\cdot
$$
\begin {equation} %---------------------------------------------------
                                                          \label {fr}
\exp \{-\int d^{4}x [\frac{1}{2}(\partial _{\alpha} \phi )^{2}
+\frac{1}{2}(\partial _{i} \phi)^{2} +V(\phi )
]\} d\phi .
\end   {equation} %---------------------------------------------------
Performing  in (\ref {fr}) the rescaling
\begin {equation} %---------------------------------------------------
                                                          \label {as2}
\phi (\lambda y,z)=\tilde {\phi }(y,z)
\end   {equation} %---------------------------------------------------
and the change of variables in the action $y\to \lambda y$ one gets
$$G_{n}(\{\lambda y_{i},z_{i}\})
=\int \tilde {\phi} (y_{1},{z}_{1})
\tilde {\phi} (y_{2},{z}_{2})...
\tilde {\phi} ( y_{n},{z}_{n})\cdot
$$
\begin {equation} %---------------------------------------------------
                                                          \label {as3}
\exp \{-\int d^{4}x [\frac{1}{2}(\partial _{\alpha} \tilde {\phi })^{2}
+\frac{1}{2}\lambda ^{2}(\partial _{i} \tilde {\phi})^{2}
 +\lambda ^{2}V(\tilde {\phi }])
\} d\tilde {\phi} .
\end {equation} %---------------------------------------------------
The anisotropic Green functions ${\cal G}_{4}(
\{\lambda y_{A\alpha},z_{Ai}\})$, $A=1,2,...n$ with small longitudinal
coordinates are related with
the  Green function
 with large longitudinal
components of momenta
${\cal G}_{n}(
\{\frac{\tilde {p}_{A\alpha}}{\lambda},p_{Ai}\})$
$$ G_{n}( \{\frac{\tilde {p}_{A\alpha}}{\lambda},p_{Ai}\}=
\int \prod _{A=1}^{n}dx_{A}e^{i\sum _{A}\frac{\tilde {p}_{A\alpha}}{\lambda}
y_{A\alpha}+p_{Ai}z_{Ai}}
G_{n}(\{ y_{A\alpha}, z_{Ai}\}=
$$
\begin {equation} %---------------------------------------------------
                                                          \label {mc}
(\lambda)^{n}
\int \prod _{A=1}^{n}dy'_{A\alpha}dz_{Ai}
e^{i\sum _{A}\tilde {p}_{A\alpha}
y_{A\alpha}+p_{Ai}z_{Ai}}
{\cal G}_{n}(\{\lambda  y'_{A\alpha}, z_{Ai}\}
\end   {equation} %---------------------------------------------------
Therefore 1PI Green functions with large longitudinal components
in momentum space
can be expressed in terms of Green functions with rescaled
longitudinal components
\begin {equation} %---------------------------------------------------
                                                          \label {gr}
G^{IPI}(\{ \frac{p_{\alpha}}{\lambda},p_{i} \})=
\frac{1}{\lambda ^{2}}
G_{\cal L_{\lambda}}^{IPI}(\{p_{\alpha},p_{i} \})
\end {equation}
 calculated in the theory with the Lagrangian
\begin {equation} %---------------------------------------------------
                                                          \label {rea}
{\cal L_{\lambda}}=\frac{1}{2}(\partial _{\alpha}\tilde { \phi})^{2}
+\frac{\lambda ^{2}}{2}(\partial _{i}\tilde { \phi})^{2} +\lambda ^{2}
V(\tilde {\phi})
\end   {equation} %---------------------------------------------------
The extra term $1/\lambda ^{2}$  in (\ref {gr})
comes from $\delta$-function
describing the momentum conservation.

Now we consider the asymptotics of the Green functions for the
 theory with
the action (\ref {rea}) when $\lambda \to 0$. Let us start with
 the free
action.

\subsubsection{Free Propagator}
Consider the action
\begin {equation} %--------------------------------------------------
        \label {free}
I_{\lambda}=\int d^{4}x [\frac{1}{2}(\partial _{\alpha} \phi )^{2}
+\frac{1}{2}\lambda ^{2}(\partial _{i} \phi)^{2}]
\end   {equation} %---------------------------------------------------
One could think that an effective action for $\lambda =0$ is just
$$
I=\int d^{4}x \frac{1}{2}(\partial _{\alpha} \phi )^{2}
$$
However we will easily see that in fact the effective action
contains another term which can be interpreted as a conformal theory
in momentum space.
The free propagator for the action (\ref {free}) has the form
\begin {equation} %---------------------------------------------------
                                                          \label {asf}
G_ {\lambda }(y,z)\equiv
\frac{1}{(2\pi )^{4}}\int \frac{e^{ikx}}
{k_{\alpha}^{2}+\lambda ^{2}k_{i}^{2}}dk
=\frac{1}{(2\pi )^{2}}\frac{1}{\lambda ^{2} y^{2}+z^{2}}
\end   {equation} %---------------------------------------------------
and is related with the standard propagator as
\begin {equation} %---------------------------------------------------
                                                          \label {pr}
G_{\lambda }(y,z)=
G(\lambda y,z)
\end   {equation} %---------------------------------------------------

Let us examine the limit of (\ref {asf}) for $\lambda \to 0$ in the sense
of theory of  distributions, i.e.
 consider the asymptotic behaviour of the integral
\begin {equation} %---------------------------------------------------
                                                          \label {asd}
( G_{\lambda},f)=\frac{1}{(2\pi )^{2}}\int d^{2}yd^{2}z
\frac{f (y,z)}{\lambda ^{2}y^{2}+ z^{2}}
\end   {equation} %---------------------------------------------------
 when $\lambda \to 0$. Here $f(y,z)$ is a test function .
We cannot simply remove $\lambda ^{2}$ from denominator since the integral
over $y$-variables diverges at $z=0$. One  subtracts  this divergence.
One has
$$
(2\pi )^{2}( G_{\lambda},f)=\int d^{2}y\int_{|z|\leq 1} d^{2}z
\frac{f (y,z)-f (y,0)}{\lambda ^{2} y^{2}+z^{2}}+
$$
\begin {equation} %---------------------------------------------------
                                                          \label {so}
\int d^{2}y\int _{|z|> 1}d^{2}z
\frac{f (y,z)}{\lambda ^{2} y^{2}+z^{2}}+\int d^{2}y\int _{|z|\leq 1}d^{2}z
\frac{f (y,0)}{\lambda ^{2} y^{2}+z^{2}}
\end   {equation} %---------------------------------------------------
 One takes $\lambda =0$ in the two first integrals
  and the answer
can be regarded as a regularized version of $1/z^{2}$.
The third integral can be calculated explicitly. One
  gets
\begin {equation} %---------------------------------------------------
                                                          \label {lex}
 G(\lambda y,z)=
\frac{1}{4\pi}  \delta ^{(2)}(z)\ln \frac{1}{\lambda ^{2}y ^{2}}
+\frac{1}{4\pi^{2}}\frac{1}{z^{2}}
+ o(1)
\end   {equation} %---------------------------------------------------
Here $$\frac{1}{z^{2}}\equiv {\mbox  Reg}\frac{1}{z^{2}},$$
\begin {equation} %---------------------------------------------------
                                                          \label {1z}
(\frac{1}{z^{2}}, f)=\int d^{2}y\int_{|z|\leq 1} d^{2}z
\frac{f (y,z)-f (y,0)}{z^{2}}+
\int d^{2}y\int _{|z|> 1}d^{2}z
\frac{f (y,z)}{z^{2}}
\end   {equation} %---------------------------------------------------

One can interprete the formula (\ref {lex}) by saying that one has two
2 dim conformal field theories here. The first one corresponds to
the first term in (\ref {lex}) and it is the standard conformal theory
but only living in 4 dim space that gives the
factor $\delta ^{(2)}(z)$. And one can interprete the second term in
(\ref {lex}) as the propagator for another conformal theory but
 living now in momentum space, i.e. we interprete $z$-coordinates as
momenta.

For  the Fourier transformation one has
\begin {equation} %---------------------------------------------------
                                                          \label {flex}
\tilde { G}_{\lambda}(k_{\alpha},k_{i})=
\frac{1}{k_{\alpha}^{2}+\lambda ^{2}k_{i}^{2}}=
\pi \delta ^{(2)}(k_{\alpha})\ln \frac{1}{\lambda ^{2}k_{i}^{2}}
+\frac{1}{(k_{\alpha})^{2}}+
+o(1),
\end   {equation} %---------------------------------------------------
$\alpha=1,2,~i=3,4.$
 Here $\frac{1}{(k_{\alpha})^{2}}=\mbox {Reg}
\frac{1}{(k_{\alpha})^{2}}$
is understood in the sense of equation (\ref {1z}).
Integrating the R.H.S. of (\ref {flex}) over $k_{\alpha}$ and
$k_{i}$ by using the relations
 \cite {Vl}
\begin {equation} %---------------------------------------------------
                                                          \label {ft2}
\int e^{ipy}\frac{d^{2}p}{p^{2}}=\pi \ln \frac{1}{y^{2}}-2\pi C_{0},
{}~~C_{0}=\int _{0}^{1}\frac{1-J_{0}(u)}{u}du-\int _{1}^{\infty}
\frac{J_{0}(u)du}{u},
\end   {equation} %---------------------------------------------------
\begin {equation} %---------------------------------------------------
                                                          \label {ft1}
\pi \int e^{ikz}\ln \frac{1}{k^{2}}d^{2}k=\frac{(2\pi)^{2}}{z^{2}}+
(2\pi)^{3}C_{0}\delta (z)
\end   {equation} %---------------------------------------------------
where $J_{0}$ is the Bessel function, we get the R.H.S. of (\ref {lex}).

For the massive case one has
\begin {equation} %---------------------------------------------------
                                                          \label {mc'}
\tilde { G}_{\lambda ,m}(k_{\alpha},k_{i})=
\frac{1}{k_{\alpha}^{2}+\lambda ^{2}k_{i}^{2}+\lambda ^{2}m^{2}}=
\pi \delta ^{(2)}(k_{\alpha})\ln \frac{1}{\lambda ^{2}(k_{i}^{2}
+m^{2})}
+ \frac{1}{k_{\alpha}^{2}}
+o(1),
\end   {equation} %----------

\subsubsection{Free generating functional}
Let us consider a  generating functional
for free Green functions for large longitudinal momenta,

\begin {equation} %---------------------------------------------------
                                                          \label {fgf}
Z(J_{\lambda})=\int \exp \{\int d^{4}x [\frac{1}{2}(\partial \phi)^{2}
+ J_{\lambda} (x) \phi (x)\} d\phi .
\end   {equation} %---------------------------------------------------
 We suppose that the source $J_{\lambda}(x)$ has the
 Fourier transformation
$ {J}_{\lambda }
(p_{\alpha},p_{i})=J_{1}(\lambda k_{\alpha}, k_{i})$
which nonvanish only for
 $|p_{\alpha}|\leq |p_{i}|$, $\alpha =0,1$ and $i=2,3$. One has
\begin {equation} %---------------------------------------------------
                                                          \label {fgfa}
Z(J_{\lambda})=\exp \{\frac{i}{2} \int d^{4}k J_{\lambda}(k)
\frac{1}{k^{2}}  J_{\lambda}(-k)\}=
\exp \{\frac{i}{2} \int d^{4}k J_{1}(\lambda k_{\alpha}, k_{i})
\frac{1}{k^{2}}J_{1}(\lambda k_{\alpha}, k_{i})\}
\end   {equation} %---------------------------------------------------
Performing a change of variables $\lambda k_{\alpha}=k'_{\alpha}$
one can represent the generating functional as
\begin {equation} %---------------------------------------------------
                                                          \label {ab}
Z(J_{\lambda})=\exp \{\frac{i}{2} \int d^{4}k J_{1}(k)
\frac{1}{k_{\alpha}^{2}+\lambda ^{2}k_{i}^{2}}
J_{1}(-k)\}
\end   {equation} %---------------------------------------------------
The behaviour of $Z(J_{\lambda})$ for $\lambda \to 0$
is given by
$$\exp \{\frac{i}{2} \int d^{4}k J_{1}(k)
(\frac{1}{(k_{\alpha})^{2}}+
\pi \delta ^{(2)}(k_{\alpha})\ln \frac{1}{\lambda ^{2}k_{i}^{2}}
+o(1)) J_{1}(-k)\}$$.
The contribution of the term with $\delta ^{(2)}$-function
dissapears due to the assumption about the support
of the current $J_{1}(k)$ and  we can say that the effective
action producing this generating functional
is simply given by
\begin {equation} %---------------------------------------------------
                                                          \label {fea}
I=\int d^{4}x \frac{1}{2}(\partial _{\alpha} \phi )^{2}
\end   {equation} %---------------------------------------------------
 Let us stress that one gets the effective action ( \ref {fea})
only under the above assumptions on the support  of the source $J_{1}(k)$.
%______________________________________________

\subsubsection{Long lines free generating functional}

Let us consider a non-local 2-lines correlation function,
\begin {equation} %---------------------------------------------------
                                                          \label {GG}
G_{\Gamma _{1}\Gamma _{2}}=\int _{\Gamma _{1}}\int _{\Gamma _{2}}
G(x)d\Gamma _{1}d\Gamma _{2},
\end   {equation} %---------------------------------------------------
which is obtained by an integration of the
usual 2-point Green function along some lines $\Gamma _{1}$ and $\Gamma _{2}$.
For example, for  the lines $\Gamma _{0}$ and $\Gamma _{1}$
which correspondes respectively to $y_{1}=0, z_{i}=0$
and $y_{0}=y_{00}, z_{i}=z_{i0}$ (see fig.1) one has

\begin{figure}
\setlength{\unitlength}{0.75cm}
\begin{center}
\begin{picture}(9,7)(-1,-1)
 \put(-2.0,0.0){\vector(1,0){9.0}}
 \put(0.0,-2.0){\vector(0,1){7.0}}
\put(-4,-2.0){\vector(2,1){11.0}}
\thinlines
\put(-2.0,2.0){\line(1,0){7.0}}
\put(-4.0,-2.0){\line(2,1){7.0}}
\put(3.0,2.5){$\Gamma _{1}$}
\put(2.0,0.5){$\Gamma _{0}$}
\put(0.2,2.5){$x_{i0}$}
\put(0.5,4.5){$i$}
\put(7.0,0.2){$0$}
\put(6.5,3.5){$1$}
  \end{picture}
\end{center}
\caption{Lines $\Gamma _{0}$ and $\Gamma _{1}$ }\label{f1}
\end{figure}
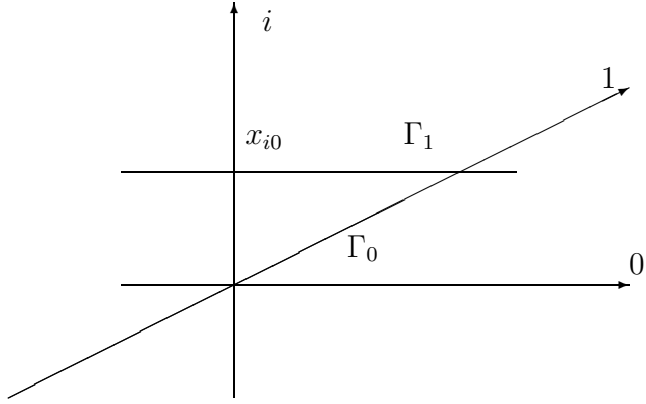

\begin {equation} %---------------------------------------------------
                                                          \label {GG}
G_{\Gamma _{0}\Gamma _{1}}(z_{i0})=\int dy_{0}\int dy_{1}
G(y_{0},y_{1},z_{i0})
\end   {equation} %---------------------------------------------------
We assume an infrared regularization so that the integration in (\ref {GG})
is performed from $-L$
to $L$. We are interested in infrared asymptotics  of this correlator
 when $L\to \infty$.
Performing the following change of variables
\begin {equation} %---------------------------------------------------
                                                          \label {Lc}
y_{\alpha}=Ly'_{\alpha}
\end   {equation} %---------------------------------------------------
we get
\begin {equation} %---------------------------------------------------
                                                          \label {icv}
G_{\Gamma _{0}\Gamma _{1}}(z_{i0})=L^{2}\int _{-1}^{1}dy'_{0}\int _{-1}^{1}
dy'_{1}G(Ly'_{0},Ly'_{1},z_{i0})=\int _{-1}^{1}dy'_{0}\int _{-1}^{1}
dy'_{1}G(y'_{0},y'_{1},\frac{z_{i0}}{L})
\end   {equation} %---------------------------------------------------
 Therefore the infrared asymptotics of (\ref {GG})
is given by the following formula
$$G_{\Gamma _{0}\Gamma _{1}}(z_{i0})=\int _{-1}^{1}dy'_{0}\int _{-1}^{1}
dy'_{1}[
\frac{1}{4\pi^{2}}\frac{1}{y'^{2}}
+ \frac{1}{4\pi}\delta ^{(2)}(y')\ln \frac{CL^{2}}{z ^{2}}
 + o(1)] =
$$
 \begin {equation} %---------------------------------------------------
                                                          \label {iicv}
\frac{1}{4\pi}\ln \frac{CL^{2}}{z ^{2}}+const .
\end   {equation} %---------------------------------------------------

The Fourier transformation of the leading term of this asymptotics
is given according to (\ref {ft2}) and (\ref {ft1}) by
\begin {equation} %---------------------------------------------------
                                                          \label {1}
\tilde {G}_{\Gamma _{0}\Gamma _{1}}(k_{i})= \int d^{2}z~e^{ik_{i}z_{i}}
G_{\Gamma _{0}\Gamma _{1}}(z_{i0})=
\end   {equation} %---------------------------------------------------
$$
(\frac{(2\pi)}{4\pi}\ln CL^{2}+8\pi ^{2}C_{0})
\delta ^{2}(k_{i})+
\frac{4\pi}{k_{i}^{2}}
$$
This correlator for $k_{i}\neq 0$ can be regarded as correlator of 2d
effective theory with a simple effective action
\begin {equation} %---------------------------------------------------
                                                          \label {ec}
S=\int d^{2}z(\partial _{i}\phi (z))^{2}
\end   {equation} %---------------------------------------------------

\section{
 Anisotropic Asymptotics and Regge Regime }
\setcounter{equation}{0}
 Let us show that
(2+2) anisotropic asymptotics for 4-points functions
are related with the Regge regime
$(s>>t)$ of the scattering amplitudes.

 Consider the 4-point 1PI Green function $\tilde {G}(k_{1},k_{2},
k_{3},k_{4})$
on the mass-shell, $k_{i}^{2}=m^{2}$.
One can choose a coordinate frame so that momenta of
particles are $ k_{1}=p_{1}+q/2$,  $k_{2}=p_{2}-q/2$,
$k_{3}= p_{2}+q/2$ and $k_{4}=p_{1}-q/2$ (see fig.2)
with $p_{1}$, $p_{2}$ and $q$ given by
\begin {eqnarray} %---------------------------------------------------
         p_{1}=(\frac{1}{2}\sqrt {s},~
         \frac{1}{2}\sqrt {-u},~0,~0),
                                                          \label {cf1}
\\
         p_{2}=(\frac{1}{2}\sqrt {s},
         -\frac{1}{2}\sqrt {-u},~0,~0)
                                                          \label {cf2}
\end   {eqnarray} %--------------------------------------------------
\begin {equation} %---------------------------------------------------
         q=(0,0,q_{i}), ~q_{i}=(q_{2},q_{3}),
                                                          \label {cf3}
\end   {equation} %---------------------------------------------------
   $s,t$ and $u$ are the Mandelstam variables and $s+u+t=4m^2$.
    \begin{figure}
\setlength{\unitlength}{0.5cm}
\begin{center}
\begin{picture}(15,7)(-5,-5)
 \put(2.5,0.0){\oval(5.0,3.0)}
\multiput(-1.5,-3.0)(6.0,4.0){2}{\line(1,1){2.0}}
\multiput(-1.5,-3.0)(6.5,4.5){2}{\vector(1,1){1.0}}
\multiput(-1.5,3.5)(6.2,-4.2){2}{\line(1,-1){2.2}}
\multiput(-1.5,3.5)(6.5,-4.5){2}{\vector(1,-1){1.0}}
 \put(-4.5,-3.5){$p_{1}+q/2$}
 \put(-4.5,2.5){$p_{2}-q/2$}
\put(6.9,-3.5){$p_{1}-q/2$}
\put(6.9,3.5){$p_{2}+q/2$}
\end{picture}
\end{center}
\caption{4-point diagram }\label{f2}
\end{figure}
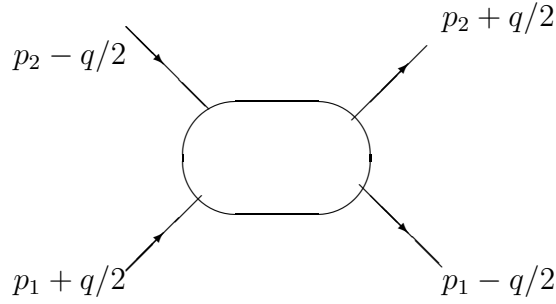

The regime  $s>>t$ corresponds to
$$s=\frac{\tilde {s}}{\lambda ^{2}} ~~\mbox{
with}~~ \tilde {s}, t~ \mbox{fixed and }~ \lambda \to 0. $$
Since in this regime $u$ also can be represent as
$u=\frac{\tilde {u}}{\lambda ^{2}} $  with $\tilde {u}$ fixed,
the longitudinal vectors $p_{1}$ and $p_{2}$ have the asymptotic
form $$p_{1}=
\frac{\tilde {p_{1}}}{\lambda },~
p_{2}=\frac{\tilde {p_{2}}}{\lambda },
{}~~\mbox{
with}~~ \tilde {p_{1}},~\tilde {p_{2}} \mbox{~~fixed and }~ \lambda \to 0$$
Therefore in the chosen coordinate frame  the regime with
$s=\frac{\tilde {s}}{\lambda ^{2}}$, $\lambda \to 0$ and  $\tilde {s} $
and $t$ fixed is described by the
1PI 4-point Green function $G_{4}(
(\frac{\tilde {p}_{1}}{\lambda},\frac{q}{2}),
(\frac{\tilde {p}_{2}}{\lambda},-\frac{q}{2}),
(\frac{\tilde {p}_{2}}{\lambda},\frac{q}{2}),
(\frac{\tilde {p}_{1}}{\lambda},-\frac{q}{2}))$
with large longitudinal coordinates in the momentum space.

\subsection{ 4-point tree diagramm}

Let us consider the diagram
  in fig.3
    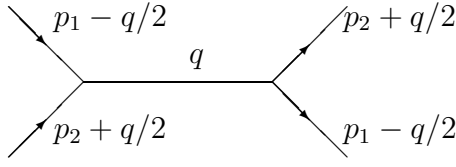
\begin{figure}
\setlength{\unitlength}{0.5cm}
\begin{center}
\begin{picture}(9,7)(-1,-1)
 \put(0.0,0.0){\line(1,0){5.0}}
 \multiput(-2.0,-2.0)(7.0,2.0){2}{\line(1,1){2.0}}
\multiput(-2.0,-2.0)(7.0,2.0){2}{\vector(1,1){1.0}}
\multiput(-2.0,2.0)(7.0,-2.0){2}{\line(1,-1){2.0}}
\multiput(-2.0,2.0)(7.0,-2.0){2}{\vector(1,-1){1.0}}
 \put(-0.8,-1.5){$p_{2}+q/2$}
 \put(-0.8,1.5){$p_{1}-q/2$}
 \put(2.8,0.5){$q$}
\put(6.9,-1.5){$p_{1}-q/2$}
\put(6.9,1.5){$p_{2}+q/2$}
\end{picture}
\end{center}
\caption{4-point tree diagram for scalar theory }\label{f3}
\end{figure}
Assume that the c.m. coordinate frame is chosen as in fig.2
  with momenta
(\ref {cf1})-(\ref {cf2}). In this frame the 1PI 4-point Green function $G_{4}(
(\frac{\tilde {p}_{1}}{\lambda},\frac{q}{2}),
(\frac{\tilde {p}_{2}}{\lambda},-\frac{q}{2}),
(\frac{\tilde {p}_{2}}{\lambda},\frac{q}{2}),
(\frac{\tilde {p}_{1}}{\lambda},-\frac{q}{2}))$
corresponding to the tree diagram (fig.2)
according
 to (\ref {gr}) is equal $\frac{1}{q^{2}_{i}},$
i.e. we get the answer which can be described by the effective action
(\ref {ec}) which is the effective action for the special non-local
correlators (\ref {GG}). Assuming that the lines $\Gamma _{1}$
and $\Gamma _{2}$ go along two light-cone lines $\Gamma _{+}$ and
$\Gamma _{-}$  we get an interpretation of $G_{\Gamma _{+}
\Gamma _{-}}$ as a scattering amplitude of two ultrarelativistic particles.

\subsection{4-point one-loop}
Now we are going to consider loop corrections.
The high-energy asymptotics of 1-loop diagram is well known \cite {Wu}
\begin {equation} %---------------------------------------------------
                                                          \label {1l}
G^{1PI}_{box}(s,t)
\rr{\sim}
g^{4}\frac{1}{2s}[\ln s-i\pi]J(q)
\end   {equation} %---------------------------------------------------
where
\begin {equation} %---------------------------------------------------
                                                          \label {Jq}
J(q)= \frac{1}{2}?\int d^{2}k_{i}
\frac{1}{k_{i}^{2}+m^{2}}\cdot
\frac{1}{(k_{i}-q_{i})^{2}+m^{2}}
\end   {equation} %---------------------------------------------------

One can see
 immediately that the same answer describes the
asymptotic behaviour for $\lambda \to 0$ of the 1-box diagramm corresponding to
the
rescaled action (\ref {rea}). Indeed the  asymptotics of the  integral
$$
G^{1PI}_{box}(\frac{\tilde {p}_{1\alpha}}{\lambda} ,
\frac{\tilde {p}_{2\alpha}}{\lambda},q )=
\frac{\lambda ^{6} g^{4}}{2}\int dk_{+}dk_{-}d^{2}k_{i}
[k_{+}k_{-}
-\lambda ^{2}k_{i}^{2}-\lambda ^{2}m^{2}+i\epsilon ]^{-1}
$$
$$
[k_{+}k_{-}
-\lambda ^{2}(k_{i}-q_{i})^{2}-\lambda ^{2}m^{2}+i\epsilon]^{-1}
[k_{-}(k_{+}-\sqrt {2\tilde{s}})
-\lambda ^{2}(k_{i}-q_{i}/2)^{2}-\lambda ^{2}m^{2}+i\epsilon]^{-1}$$
\begin {equation} %---------------------------------------------------
                                                          \label {as8}
[k_{+}(k_{-}+\sqrt {2\tilde{s}})
-\lambda ^{2}(k_{i}-q_{i}/2)^{2}-\lambda ^{2}m^{2}+i\epsilon]^{-1}
\end   {equation} %---------------------------------------------------
$\tilde{s}=(\tilde {p}_{1\alpha}+\tilde {p}_{2\alpha})^{2}$,
is
\begin {equation} %---------------------------------------------------
                                                          \label {as9}
G^{1PI}_{box}(\frac{\tilde {p}_{1\alpha}}{\lambda} ,
\frac{\tilde {p}_{2\alpha}}{\lambda},q )\osim{\sim}
g^{4}\frac{\alpha ^{2}}{2s}[\ln s/\alpha ^{2}-i\pi]J(q).
\end   {equation} %---------------------------------------------------

Note that (\ref {as9}) takes place only for massive theory. The asymptotics
of box diagram with massless internal lines and massive external line
is
\begin {equation} %---------------------------------------------------
                                                          \label {asml}
G^{1PI}_{box}(\frac{\tilde {p}_{1\alpha}}{\lambda} ,
\frac{\tilde {p}_{2\alpha}}{\lambda},q )\osim{\sim}
\frac{i\pi^{2}g^{4}\lambda ^{2}}{\tilde {s}t}(\ln \frac{m^{4}\lambda ^{2}}
{\tilde {s}t})^{2}
\end   {equation} %---------------------------------------------------
that is in the agreement with exact answer for the massless box diagram
 \cite {NU}.

 \section{Anisotropic asymptotics for gauge theories}
\setcounter{equation}{0}
Let us now consider the anisotropic
 asymptotics of  correlation
functions for the gauge theory
\begin {equation} %---------------------------------------------------
                                                          \label {tr}
G_{n}(\{\lambda y_{j},z_{j} \})=
<A_{\mu _{1}}(\lambda y_{1},z_{1})
...
A_{\mu _{n}}(\lambda y_{n},z_{n})> \mbox{~for }
\lambda \to 0.
\end   {equation} %---------------------------------------------------
$<...>$ means the average
\begin {equation} %---------------------------------------------------
                                                          \label {FI}
<A_{\mu _{1}}(\lambda y_{1},z_{1})...
A_{\mu _{n}}(\lambda y_{n},z_{n})>=
\end   {equation} %---------------------------------------------------

$$
\int
A_{\mu _{1}}(\lambda y_{1},z_{1})
...A_{\mu _{n}}(\lambda y_{n},z_{n})
\exp i\int d^{4}x \{\tr [F_{\mu\nu}^{2}
+\frac{1}{2\beta}(\partial _{\mu} A_{\mu})]+\bar {c}Mc\}
dAd\bar {c}dc,$$
where
$F_{\mu\nu}$ is the field strength,
$ F_{\mu\nu} =
\partial _\mu A_\nu - \partial_\nu A_\mu + [A_\mu,A_\nu]$ ,
$A_\mu = A_\mu^a \tau^a $,
$g$ is the coupling constant and
$\tau^a$ are the generators of the Lie algebra of the
gauge group $G = SU(N)$, $M=\partial _{\mu} D_{\mu},$
$D_{\mu}=\partial _{\mu} +A_{\mu}$.
Let us note that the r.h.s. of (\ref {FI}) has a rather formal meaning.
One has to assume some regularization procedure and
moreover the functional integral
 being understood  in the perturbation theory
can be used to calculate correlation functions only for
short distances.
Let us ignore for a moment these subtleties and
  perform an estimation
of the asymptotics of r.h.s. of (\ref {FI}) by using the anisotropic
 dimensional
analysis. Performing the change of variables
\begin {equation} %---------------------------------------------------
                                                          \label {cv}
A_{\alpha}(\lambda y,z)=\frac{1}{\lambda }\tilde {A}_{\alpha}(y,z),~~
A_{i}(\lambda y,z)=\tilde {A}_{i}(y,z)
\end   {equation} %---------------------------------------------------
in the path integral (\ref {FI}) and also the
change of variables in the integral over $y$-space
we see that
 the correlation functions (\ref {tr}) can be computed with
the rescaled Yang-Mills action, i.e.
\begin {equation} %---------------------------------------------------
                                                          \label {rcf}
<A_{\mu _{1}}(\lambda y_{1},z_{1})
...
A_{\mu _{n}}(\lambda y_{n},z_{n})>=
(\frac{1}{\lambda} )^{n}<A_{\mu _{1}}(y_{1},z_{1})
...
A_{\mu _{n}}(y_{n},z_{n})>_{S_{\lambda}},
\end   {equation} %---------------------------------------------------

$$<A_{\mu _{1}}( y_{1},z_{1})
...
A_{\mu _{n}}( y_{n},z_{n})>_{S_{\lambda}}=
$$
\begin {equation} %---------------------------------------------------
                                                          \label {rcf'}
\int A_{\mu _{1}}(y_{1},z_{1})
...
A_{\mu _{n}}(y_{n},z_{n})
\exp \{i S_{\lambda} \}~dA,
\end   {equation} %---------------------------------------------------
where $\mu _{k_{j}}=0,1$ for $j=1,... m$ and $\mu _{k_{j}}=2,3$ for
 $j=m+1,...n$ and
\begin {equation} %---------------------------------------------------
                                                          \label {ra}
S_{\lambda} = \frac{1}{4\lambda ^{2}g^{2}} \int d^{4}x \tr (
F_{\alpha \beta}F^{\alpha \beta})
+\frac{1}{2g^{2} }\int d^{4}x\tr(F_{\alpha  j}F^{\alpha  j})+
\frac{\lambda ^{2}}{4g^{2} }\int d^{4}x\tr(F_{ij}F^{ij})
\end   {equation} %---------------------------------------------------
$$+ \mbox{gauge fixing terms}
$$
The action (\ref {ra}) has the form of Verlinde's \cite {VV} rescaled
action. Let us discuss the limit $\lambda \to 0$.
Note that the rescaling
(\ref {cv})  for
 the Wilson loop operator $W(\Gamma _{l})=$
$P\exp \int _{\Gamma _{l}}A_{\alpha}dx^{\alpha}$ for  loops
 belonging to the longitudinal plane gives the relation
\begin {equation} %---------------------------------------------------
                                                          \label {wl}
<W(\Gamma ^{1}_{l})W(\Gamma ^{2}_{l})>
=<W(\Gamma '^{1}_{l})W(\Gamma '^{2}_{l})>_{S'}
\end   {equation} %---------------------------------------------------
where $\Gamma '^{i}$ is the rescaled loop.  In particular for the infinite
long lines $\Gamma _{+}$, $\Gamma _{-}$, one
 gets at the formal level the
relation
\begin {equation} %---------------------------------------------------
                                                          \label {wlpm}
<W(\Gamma _{+})W(\Gamma _{-})>
=<W(\Gamma _{+})W(\Gamma _{-})>_{S'}.
\end   {equation} %---------------------------------------------------

Now
 one can try
to consider the parameter $\lambda$ like a parameter in
the quassiclassical expansion. To make sense to such an expansion one has to
 use
 a regularization. A more suitable regularization is the lattice
regularization.
Since we intend to study the theory in the region of  small longitudinal
coordinates and large transversal
coordinates and we want to take into account fluctuations in different
direction with an approximately equal precision it is relevant to assume
that the lattice spacing in the longitudinal and transversal
directions are different so that there are
equal number of points in the different directions (see fig 4) \cite {AII}.
If the number of points
in different directions is the same, i.e. $n_{0}=n_{1}=...n_{D-1}$
then we get a theory in an asymmetric space-time volume
$L_{0}L_{1}...L_{d-1}$, $L_{i}$ is a typical size in the i-direction,
with lattice spacing $a_{0},a_{1},...a_{D-1},$ such that
$L_{0}/L_{1}=a_{0}/a_{1}$,
..., $L_{0}/L_{D}=a_{0}/a_{D}$, .

A general form of the lattice
action on an asymmetric lattice with lattice spacing $a_{0},a_{1},...,a_{D-1}$
in 0,1,...D-1-directions has the form
\begin {equation}
                                                          \label {aa}
S=\frac{1}{4g^{2}}a_{0}a_{1}...a_{D-1}\sum _{x,\mu, \nu  }
\frac{1}{(a_{\mu}a_{\nu})^{2}}\tr (U(\Box _{\mu, \nu })-1),
\end   {equation}
Here $x$ are  points of the 4-dimensional lattice,
$\Box _{\mu , \nu }$  is a single plaquette
attached to the links $(x,x+\mu)$ and $(x, x+\nu)$, $U(\Box _{\mu , \nu })=
U_{x,\mu}U_{x+\mu,\nu}U^{+}_{x+\nu,\mu}U^{+}_{x,\nu}$ and link variables
$U_{x,\mu}$ are associated with the link between the lattice sites $x$ and
$x+\hat {\mu}$. $U_{x,\mu}$ belongs to a representation of
the gauge group $SU(N)$.

Since we are interested in the case when $L_{0}=L_{1}$, $L_{2}=L_{3}$ and
$L_{1}/L_{3}\to 0$ we put in (\ref {aa})  $a_{0}=a_{1}$, $a_{2}=a_{3}$
and $a_{0}=\lambda a_{3}$, so we get
\begin {equation}
                                                          \label {al}
S=\frac{1}{4\lambda ^{2}g^{2}}\sum _{x}\sum _{\alpha,\beta }
\tr (U(\Box _{\alpha,\beta })-1) +
\frac{1}{4g^{2}}\sum _{x}\sum _{\alpha, i }
\tr (U(\Box _{\alpha, i })-1)
\end   {equation}
$$+\frac{\lambda ^{2}}{4g^{2}}\sum _{x}\sum _{i,j }
\tr (U(\Box _{i,j})-1).$$
 Here $\alpha$ and $\beta $ are unit
vectors in the longitudinal direction and $i,j$ are unit vectors in
the transversal direction. We also denote the
points of 4-dimensional lattice as $x=(y,z)$, where $y$ and $z$
are the points of two two-dimensional lattices, say, y-lattice (longitudinal)
and z-lattice (transversal).
Performing the $\lambda \to 0$ limit in the lattice action (\ref {aa})
we get
\begin {equation} %---------------------------------------------------
                                                          \label {a}
S^{tr}=\frac{1}{4g^{2}}\sum _{x}\sum _{\alpha, i }
\tr (U(\Box _{\alpha, i }) -1),
\end   {equation} %---------------------------------------------------
with $U_{x,\alpha}$ being a subject of the relation
\begin {equation} %---------------------------------------------------
                                                          \label {zc}
U(\Box _{\alpha,\beta })=1.
\end   {equation} %---------------------------------------------------
Therefore $U _{x,\alpha }$ is a zero-curvature lattice gauge field
\begin {equation} %---------------------------------------------------
                                                          \label {z}
U _{x,\alpha }=V _{x}V ^{+}_{x+\alpha }.
\end   {equation} %---------------------------------------------------
Substituting (\ref {z}) in (\ref {a}) we get
\begin {equation} %---------------------------------------------------
                                                          \label {1}
S^{tr}=\frac{1}{4g^{2}}\sum _{x}\sum _{\alpha, i }
\tr (V_{x+\alpha}^{+}U_{x+\alpha,i}V_{x+\alpha +i}
V^{+}_{x+i}U^{+}_{x, i }V_{x}-1).
\end   {equation} %---------------------------------------------------
or
\begin {equation} %---------------------------------------------------
                                                          \label {A}
S^{tr}=\frac{1}{4g^{2}}\sum _{x}\sum _{\alpha, i }
\tr (\tilde {U}_{x+\alpha,i} \tilde {U}_{x,i}-1)
\end   {equation} %---------------------------------------------------
where
\begin {equation} %---------------------------------------------------
                                                          \label {V}
\tilde {U}_{x}=V^{+}_{x+i}U^{+}_{x, i }V_{x}.
\end   {equation} %---------------------------------------------------
The action (\ref {A}) is ultra-local in the transverse z-direction
and
 it is the lattice chiral field action in the longitudinal y-direction.
In the formal
continuum limit $a_{0}\to 0$ we get
\begin {equation} %---------------------------------------------------
                                                          \label {cl}
S^{tr}_{c,l}=\frac{1}{4g^{2}}\int dy^{0}dy^{1}\sum _{z, i, \alpha }
\tr[\partial _{\alpha}\tilde {U}_{z,i}(y)\partial _{\alpha}
\tilde {U}^{+}_{z,i}(y)]
\end   {equation} %---------------------------------------------------
with summation over the repeating indices $\alpha =0,1$ and
$\tilde {U}_{z,i}(y)=V^{+}_{z+i}(y)U^{+}_{z, i }(y)V_{z}(y)$.
To get its pure continuous version one has
also to make in (\ref {cl}) and (\ref {V}) a formal limit $a_{t}\to 0$
under
 the assumption $U_{x,i}=exp(a_{t}A_{i})$
\begin {equation} %---------------------------------------------------
                                                          \label {cc}
S^{tr}_{c,c}=\frac{1}{4g^{2}}\int dy^{0}dy^{1}d^{2}z_{i} \sum _{\alpha ,i }
\tr[\partial _{\alpha}\tilde {A}_{i}\partial _{\alpha}\tilde {A}_{i}]
\end   {equation} %---------------------------------------------------
where
\begin {equation} %---------------------------------------------------
                                                          \label {VV}
\tilde {A}_{i}=V^{+}D_{i}V,~~ D_{i}=\partial _{i}+A_{i},
\end   {equation} %---------------------------------------------------
 \begin{figure}
\setlength{\unitlength}{0.5cm}
\begin{center}
\begin{picture}(5,5)(-1,-2)
 \multiput(0.0,0.0)(0.5,0.0){5}{\line(0,1){4.0}}
 \multiput(0.0,0.0)(0.0,1.0){5}{\line(1,0){2.0}}
 \put(-1.0,1.0){$a_{2}$}
\put(0.0,-1.0){$a_{1}$}
\put(2.0,-1.0){$L_{1}$}
\put(-1.0,4.0){$L_{2}$}
 \put(-2.0,-2.5){$n_{1}=n_{2},~~\frac{L_{1}}{L_{2}}=\frac{a_{1}}{a_{2}}$}
\end{picture}
\end{center}
\caption{Asymmetric lattice }\label{f1}
\end{figure}
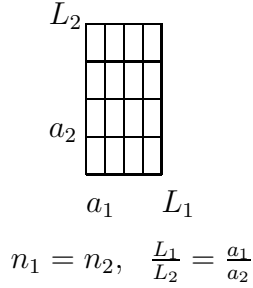

Therefore in the formal limit $a_{t} \to 0 $ the action (\ref {cl})
reproduces the  Verlinde and
Verlinde truncated action. However
non-perturbatively there is an essential
difference in the behaviour of two theories (\ref {cc}) and (\ref {cl}).

The lattice version of
 (\ref {wlpm}) for the zero curvature  longitudinal
gauge fields is simply reduced to the boundary values of the field $V_{z}(y)$
\begin {equation} %---------------------------------------------------
                                                          \label {lsf}
{\cal V}_{+}(L,0,z) =V_{z}(-L,0)V^{+}_{z}(L,0),~~
{\cal V}_{-}(0,L,z) =V_{z}(0,-L)V^{+}_{z}(0,L).
\end   {equation} %---------------------------------------------------
and the expectation value of these string operators is given by
\begin {equation} %---------------------------------------------------
                                                          \label {EA}
<{\cal V}_{+}(L,0,0){\cal V}_{-}(0,L,z)>=
\end   {equation} %---------------------------------------------------
$$\int V_{0}(-L,0)V^{+}_{0}(L,0)
V_{z}(0,-L)V^{+}_{z}(0,L)\exp S^{tr}(V,U) dV dU$$
with $S^{tr}(V,U)$ as in (\ref {A}). In the continuum limit in the
longitudinal direction we use $S^{tr}$ given by (\ref {cl}).
Our goal consists in calculation the functional integral (\ref {EA})
over gauge fields U to get an effective action describing an interaction
of fields $V_{z}(0,\pm L)$ $V_{z}(\pm L,0)$.
The action (\ref {A}) is the action for two-dimensional lattice
non-linear $\sigma$-model in finite volume.
Transversal dynamics arises from boundary effects for two-dimensional
(longitudinal) non-linear $\sigma$-model.
Therefore to get an effective action
describing the transversal dynamics we have to calculate the Schrodinger
functional for two-dimensional non-linear $\sigma$-model  \cite {LuSy}.
An exact solution of this problem is unknown. An approximate effective action
is obtained in \cite {AII} under the assumption that the  $\sigma$-model
has massive excitations.

$$~$$
{\bf ACKNOWLEDGMENT}
$$~$$
This work has been supported in part by
International Science Foundation under the grant M1L000.
 We are grateful to L.Accardi, J.Bartels, Yu.N.Drozzinov, M.Caselle,
L.D.Faddeev, F.Gliozzi, M.Lusher, A.M.Polyakov, E.Verlinde,
V.S.Vladimirov and B.I.Zavialov
for
useful discussions.
$$~$$

\end{document}